\documentclass[twocolumn,printnumbers,amsmath,amssymb,showpacs,prl]{revtex4}
\usepackage{graphicx}
\usepackage{color}

\begin{document}
\title{Structural heterogeneity and its role in determining properties of disordered solids}

\date{\today}

\author{Hua Tong}
\author{Ning Xu$^*$}

\affiliation{CAS Key Laboratory of Soft Matter Chemistry, Hefei National Laboratory for Physical Sciences at the Microscale, and Department of Physics, University of Science and Technology of China, Hefei 230026, People's Republic of China.}

\begin{abstract}
We construct a new order parameter from the normal modes of vibration, based on the consideration of energy equipartition, to quantify the structural heterogeneity in disordered solids.  The order parameter exhibits strong spatial correlations with low-temperature single particle dynamics and local structural entropy.  To characterize the role of particles with the most defective local structures identified by the order parameter, we pin them and study how properties of disordered solids respond to the pinning.  It turns out that these particles are responsible to the quasilocalized low-frequency vibration, instability, softening, and nonaffinity of disordered solids.
\end{abstract}


\maketitle

The nature of disordered solids, {\it e.g.}~glasses and sandpiles, remains elusive and a major challenge to condensed matter physics \cite{Binder,Berthier1,Berthier2}.  Compared to crystalline solids, the absence of long-range structural order makes it difficult to interpret properties of disordered solids analytically.  What makes it more difficult is the spatial heterogeneity of the structural disorder.  It has been evidenced that the heterogeneous disorder greatly contributes to abnormal properties of disordered materials, {\em e.g.}~the anomalous low-frequency excitations and consequent unusual thermal properties \cite{Boson}, heterogeneous mechanical response to perturbations \cite{Yoshimoto,Tsamados,Ellenbroek}, and dynamical heterogeneity of supercooled liquids \cite{Tanaka1,Tanaka2,Harrowell1,Harrowell2}.  Therefore, how to correctly describe the heterogeneous disorder is the key to develop the theory of disordered solids.

For crystals, it has been well-known that dislocations are triggers of the instability, which have lower bond orientational order than perfect lattice sites and can thus be easily identified.  The bond orientational order has been applied to identify defective spots in weakly disordered solids \cite{Tanaka1,Tanaka2,Tanaka3,Tanaka4}. However, this approach fails to describe the structural heterogeneity of strongly disordered solids, {\it e.g.}~systems with large particle size dispersity, in which the locally favored geometric structure is no longer a perfect crystal \cite{Tanaka2,Tanaka4}.  An alternate order parameter is thus needed to pick out ``defective" structures in disordered solids, which must capture the heterogeneous dynamics correctly and be responsible to the special properties of disordered solids.

Inspired by recent observations that low-frequency quasilocalized modes of vibration are correlated with particle rearrangements in disordered systems \cite{Harrowell2,Brito,Ghosh,Chen,Manning}, we construct an order parameter $\Psi$ at single particle level from normal modes of vibration, based on the assumption of energy equipartition.  This new order parameter is validated by showing excellent spatial correlations with low-temperature dynamics and structural entropy.  In order to figure out the role of particles with the largest $\Psi$, {\it i.e.}~particles with the most defective local structures, we measure the system response to the pinning of these particles.  Interestingly, the pinning remarkably eliminates the low-frequency quasilocalized modes, strengthens the system stability upon thermal excitation, suppresses the nonaffine deformation under shear or compression, and hardens the solids with higher shear and bulk moduli.  These observations reveal the key role of particles with the most defective local structures in determining special properties of disordered solids.

We study both two- (2D) and three-dimensional (3D) systems with side length $L$ and periodic boundary conditions in all the directions.  To avoid crystallization, we use a $50:50$ binary mixture of $N=1024$ spheres (disks) with equal mass $m$ and a diameter ratio $1.4$.  The interaction potential between particles $i$ and $j$ is
\begin{equation}
V(r_{ij})=\frac{\epsilon}{72}\left[ \left(\frac{\sigma_{ij}}{r_{ij}}\right)^{12}-\left(\frac{\sigma_{ij}}{r_{ij}}\right)^{6}\right]+f(r_{ij}),
\label{Eq: Potential}
\end{equation}
when $r_{ij}/\sigma_{ij}<2.5$ and zero otherwise, where $r_{ij}$ is the particle separation, $\sigma_{ij}$ is the sum of particle radii, and $f(r_{ij})$ guarantees that the potential and its first derivative are zero at $r_{ij}=2.5\sigma_{ij}$.  We set the units of length, mass, and energy to be small particle diameter $\sigma$, particle mass $m$, and characteristic energy scale $\epsilon$.  Time and temperature are in units of $\sqrt{m\sigma^2/\epsilon}$ and $\epsilon/k_B$ with $k_B$ the Boltzmann constant.  The packing fraction $\phi$ is determined from the repulsive core.  Results shown here are for $\phi_{_{3D}}=0.75$ and $\phi_{_{2D}}=0.95$.  We have verified that our major findings are general for other packing fractions and other types of interaction potential.

We generate zero temperature ($T=0$) glasses by quickly quenching high temperature states to their local potential energy minima using the fast inertial relaxation engine minimization algorithm \cite{Fire}.  The normal modes of vibration are obtained by diagonalizing the dynamical matrix using ARPACK \cite{arpack}, from which we obtain the density of states $D(\omega)=\langle \sum_j \delta(\omega-\omega_j) \rangle/ N$ and participation ratio $p(\omega)=\langle\sum_j p_j\delta(\omega-\omega_j)/\sum_j \delta(\omega-\omega_j) \rangle$, where $\omega_j$, $p_j=(\sum_{i=1}^N |\vec{e}_{j,i}|^2)^2/N\sum_{i=1}^N |\vec{e}_{j,i}|^4$, and $\vec{e}_{j,i}$ are the frequency, participation ratio, and polarization vector of particle $i$ of mode $j$, $\left< .\right>$ denotes the average over $1000$ configurations, and the sums $\sum_j$ are over all modes.

Assuming that upon excitations the vibrational energy is equally distributed to all modes, the mean square vibrational amplitude of particle $i$ is proportional to
\begin{equation}
\Psi_i = \sum_{j=1}^{dN-d}\frac{1}{\omega_j^2}|{\vec{e}}_{j,i}|^2, \label{Eq: MSVAep}
\end{equation}
where $d$ is the dimension of space.  There are $d$ zero-frequency modes due to the translational invariance imposed by periodic boundary conditions, so the total number of nontrivial vibrational modes is $dN - d$.  Particles with larger $\Psi$ tend to move more freely, in analogy with dislocations in crystals.  Therefore, we define $\Psi$ as the order parameter to characterize the heterogeneous structure of disordered solids at single particle level.

To test the effectiveness of $\Psi$, we measure the single particle dynamics via molecular dynamics simulations at $T=10^{-6}$ (much lower than the glass transition temperature), by adding thermal energy to the $T=0$ glasses.  Although systems do not relax, some less stable ones still exhibit cage jumps with occasional local particle rearrangements, which are excluded in our measurement of local dynamics to ensure that the thermalized systems stay in the same basins of attraction of the $T=0$ glasses.  Cage jumps will be discussed later.

\begin{figure}
    \includegraphics[width=0.45\textwidth]{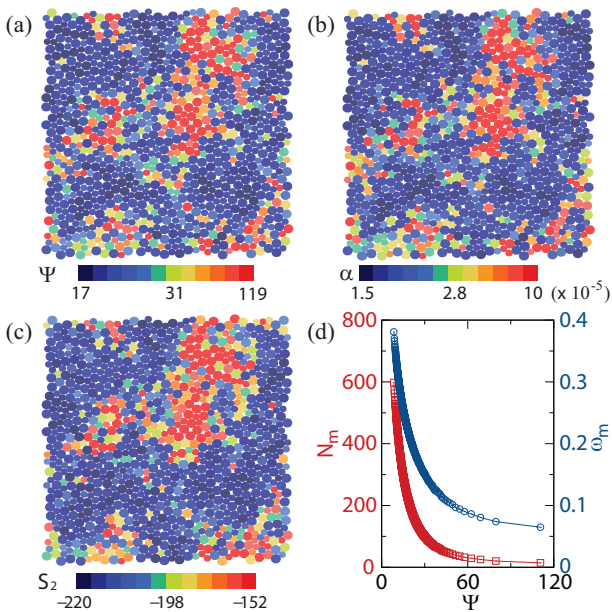}
  \caption{(color online).  Spatial distributions of (a) order parameter $\Psi$, (b) LDWF $\alpha$, and (c) LSE $S_2$ for a 2D system. (d) Order parameter $\Psi$ dependence of the number of modes $N_m$ (squares) and weighted frequency $\omega_m$ (circles) defined in the text for 3D glasses, with the solid curves to guide the eye.}
  \label{fig1}
\end{figure}

We measure the local Debye-Waller factor (LDWF) $\alpha$ and local structural entropy (LSE) $S_2$ to characterize the low-temperature dynamics.   The LDWF for particle $i$ is defined as $\alpha_i = \langle [\vec{r}_i(t)-\vec{r}_i(0)]^2\rangle_{_{t}}$, where $\vec{r}_i(t)$ is the position of particle $i$ at time $t$ and $\left< .\right>_{_t}$ denotes the time average.  For supercooled liquids or glassy states, the mean square displacement (MSD) exhibits a plateau between the short-time ballistic motion and long time diffusion.  The LDWF measures the plateau value of the MSD at single particle level and the time average is taken after the MSD has reached its plateau.  The LSE for particle $i$ is defined as $S_{2,i}=-1/2\sum_{\nu}\rho_{\nu}\int d{\vec{r}}\left\{g^{\mu\nu}_i({\vec{r}}){\rm ln}g^{\mu\nu}_i({\vec{r}})-\left[g^{\mu\nu}_i({\vec{r}})-1\right]\right\}$, where $\mu$ and $\nu$ denote the type of particles (large or small), $\rho_{\nu}$ is the number density of type $\nu$ particles, and $g^{\mu\nu}_i(\vec{r})$ is the pair correlation function between particle $i$ of type $\mu$ and the other particles of type $\nu$ and is also obtained from the time average.  It has been shown that the LDWF and LSE are correlated well with the heterogeneous dynamics \cite{Harrowell1,Tanaka2} and diffusive behaviors \cite{Tanaka1,Tanaka2,Samanta,Mittal,Private} in supercooled liquids.  Both quantities are dynamically determined and accessible to experiments of colloidal systems.

Figure~\ref{fig1} compares the spatial distributions of $\Psi$, $\alpha$, and $S_2$ for a 2D configuration.  Strong correlations between these quantities are apparent.  Particles with larger $\Psi$ also have larger values of $\alpha$ and $S_2$.  We calculate $C_{A,B}=\frac{\sum_{i=1}^N(A_i-\langle A\rangle)(B_i-\langle B \rangle)}{\sqrt{\sum_{i=1}^N(A_i-\langle A\rangle)^2}\sqrt{\sum_{i=1}^N(B_i-\langle B\rangle)^2}}$ to quantify the correlation, where $A,B=\Psi,\alpha,S_2$, and $\left< .\right>$ denotes the particle average.  $A$ and $B$ are more correlated with larger $C_{A,B}$ ($\in[0,1]$).  We find that $C_{\Psi,\alpha}=0.994\pm0.005$ ($0.985\pm0.009$) and $C_{\Psi,S_2}=0.704\pm0.046$ ($0.806\pm0.051$) in 3D (2D), indicating that $\Psi$ is indeed strongly correlated with $\alpha$ and $S_2$.

The excellent correlation between $\Psi$ and $\alpha$ is a direct consequence of energy equipartition, which just trivially verifies that at low temperatures the energy is equally distributed to all the normal modes of vibration.  In contrast, the strong correlation between $\Psi$ (or $\alpha$) and $S_2$ is not so obvious.  $S_2$ has also been proposed as the measure of local structural order \cite{Tanaka1,Tanaka2}.  The strong correlation between $\Psi$ and $S_2$ thus proves the effectiveness of $\Psi$ as the structural order parameter of disordered solids at $T=0$ in the absence of dynamics.

Recently, particles with large polarization vectors in tens of lowest-frequency modes have been used to identify defective or soft spots responsible to localized particle rearrangements under excitations \cite{Chen,Manning,Ghosh,Harrowell2}.  Next we will show that this identification of soft spots is consistent with our approach.  For particle $i$, we sort the modes in the descending order of $\frac{1}{\omega_j^2}|\vec{e}_{j,i}|^2$ ($j=1, 2, ..., dN-d$) and calculate the smallest number of modes $N_{m,i}$ satisfying $\sum^{N_{m,i}}_{j=1}\frac{1}{\omega^{\prime 2}_j}\left| \vec{e}^{~\prime}_{j,i} \right|^2 / \sum^{dN-d}_{j=1}\frac{1}{\omega^{\prime 2}_j}\left| \vec{e}^{~\prime}_{j,i} \right|^2 \ge 0.8$, where $\omega^{\prime}_j$ and $\vec{e}^{~\prime}_{j,i}$ are the frequency and polarization vector of particle $i$ of the sorted mode $j$.  Correspondingly, we calculate the weighted average frequency $\omega_{m,i}$ of these $N_{m,i}$ modes: $\omega_{m,i}=\sum_{j=1}^{N_{m,i}}\frac{1}{\omega^{\prime }_j}\left| \vec{e}^{~\prime}_{j,i} \right|^2 / \sum^{N_{m,i}}_{j=1}\frac{1}{\omega^{\prime 2}_j}\left| \vec{e}^{~\prime}_{j,i} \right|^2$.  Therefore, $N_{m,i}$ and $\omega_{m,i}$ indicate how many modes contribute to most of particle $i$'s vibration and where they are in the frequency domain.  In Fig.~\ref{fig1}(d), we plot $N_m$ and $\omega_m$ against $\Psi$.  Apparently, vibrations of particles with large $\Psi$ are contributed by a small number of modes (in the order of $10$) with low frequencies, which explains why decent spatial correlations between only tens of lowest-frequency modes and heterogeneous glassy dynamics have been observed \cite{Chen,Manning,Ghosh,Harrowell2}. Compared to previous work, the order parameter $\Psi$ defined here is more accurate to identify particles with defective local structures in disordered solids with a clear theoretical origin.

\begin{figure}[t]
    \includegraphics[width=0.45\textwidth]{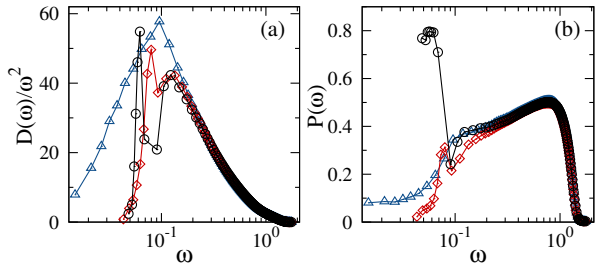}
  \caption{(color online). (a) Reduced density of states $D(\omega)/\omega^2$ and (b) participation ratio $P(\omega)$ of 3D $T=0$ glasses.  The circles, diamonds and triangles are for pinning $3\%$ particles with the largest $\Psi$, randomly pinning $3\%$ particles, and no pinning, with the solid curves to guide the eye.}
  \label{fig2}
\end{figure}

We have shown that particles with the largest $\Psi$ have the greatest potential to move.  In analogy with dislocations in crystals, these particles should be responsible to the instability and play a key role in mechanical response to perturbations.  It is then interesting to know if suppressing their motion can significantly make the solids more stable to excitations and alter their vibrational and mechanical properties.  To check it, we pin particles with the largest $\Psi$ and measure the response of solid properties.  As a comparison, we also repeat the same measurements by pinning randomly selected particles.

Normal modes of vibration are the fundamentals to understanding properties of solids and are thus our first concern.  Note that pinning particles breaks the translational invariance of periodic boundary conditions, which converts the $d$ zero-frequency modes to nontrivial modes with nonzero frequencies.  Interestingly, pinning particles with the largest $\Psi$ has distinct effects from random pinning on the behaviors of these $d$ modes.  As shown in Fig.~\ref{fig2}, although these modes constitute a peak (the one at lower frequencies) in the density of states for both pinning cases, they have much larger participation ratio $p(\omega)$ and lower frequencies for the pinning of particles with the largest $\Psi$.  This distinction strongly supports that particles with the largest $\Psi$ have the most defective local structures. Suppressing their motion makes the system effectively so uniform to long wavelengths that only weak heterogeneity is probed.  In contrast, random pinning does not have a good control of the most defective local structures, which still act as scatters and make the modes more localized.

Special vibrational features of glasses include the boson peak and low-frequency quasilocalization.  Excess number of modes beyond the Debye law, {\it i.e.}~$D(\omega)\sim \omega^{d-1}$, form a peak in $D(\omega)/\omega^{d-1}$ called boson peak, as clearly demonstrated by the triangles in Fig.~\ref{fig2}(a).  When pinning is implemented, the boson peak (second peak at higher frequencies) shifts to higher frequencies, but the shift seems independent of the pinning protocol.  In contrast, the low-frequency quasilocalization is very sensitive to the pinning protocol.  As shown in Fig.~\ref{fig2}(b), pinning particles with the largest $\Psi$ remarkably eliminates the low-frequency quasilocalized modes, while with random pinning the quasilocalized modes still survive but shift to higher frequencies.

It has been shown that the boson peak frequency and low-frequency quasilocalization affects the stability of glasses subject to excitations \cite{xu,wang,singh}: the glass is more stable with higher boson peak frequency and weaker quasilocalization [larger $p(\omega)$].  Figure~\ref{fig2} indicates that both pinning protocols stabilize the glass by increasing the boson peak frequency.  Compared to random pinning, pinning particles with the largest $\Psi$ stabilize the glass further by eliminating low-frequency quasilocalized modes, which again strongly supports that particles with the largest $\Psi$ have the most defective local structures and are in consequence responsible to the instability of disordered solids.  To verify it further, we investigate how the cage jumps mentioned earlier are affected by    pinning.

\begin{figure}[t]
    \includegraphics[width=0.45\textwidth]{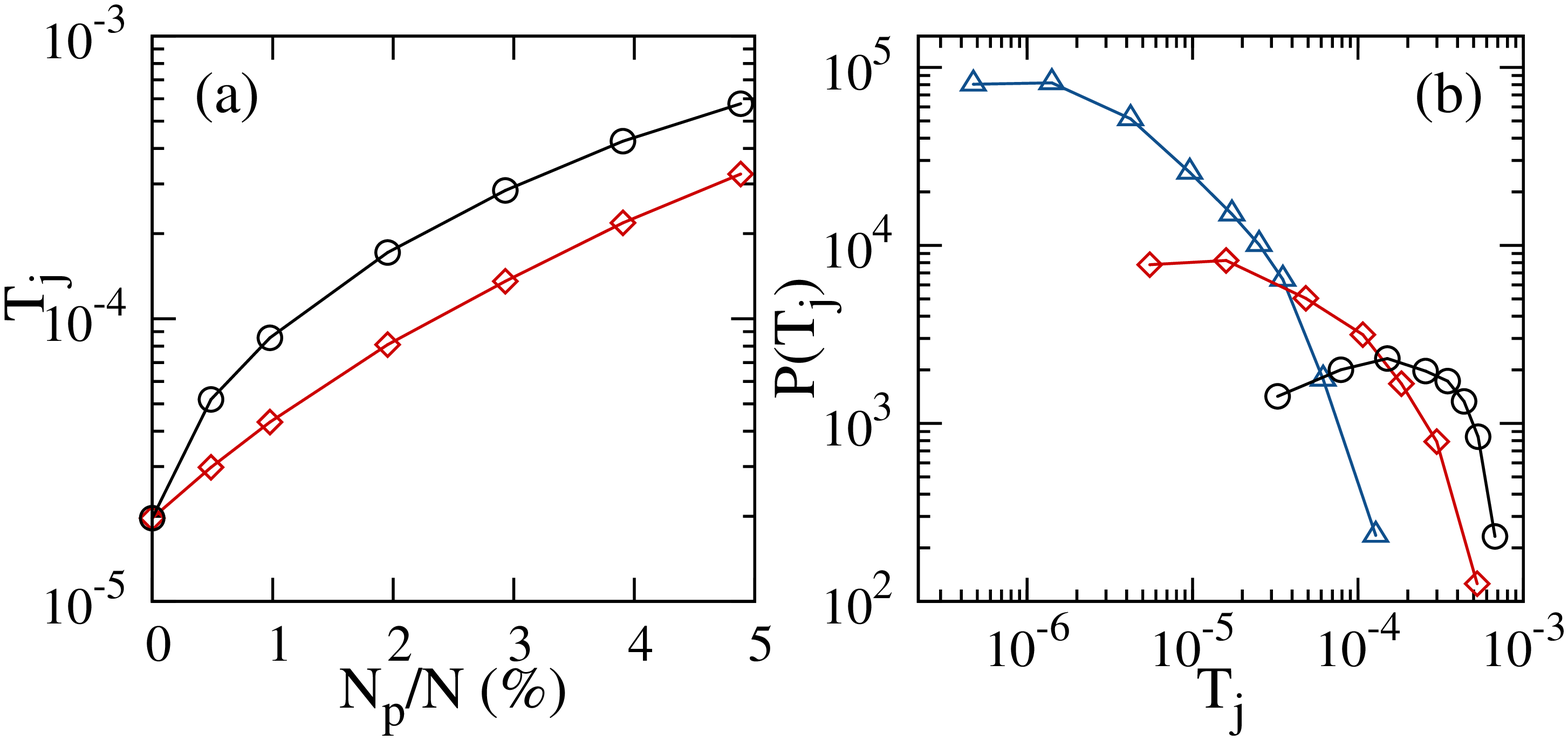}
  \caption{(color online). (a) Dependence of the cage-jump temperature $T_j$ on the fraction of pinned particles $N_p/N$ for 3D systems. (b) Distributions of $T_j$ for pinning $3\%$ particles with the largest $\Psi$ (circles), randomly pinning $3\%$ particles (diamonds), and no pinning (triangles).  The solid curves are to guide the eye.}
  \label{fig3}
\end{figure}

When a glass is heated to a temperature below the glass transition temperature, although global relaxation is inaccessible, local topological change can occasionally occur, which drives the system into a neighbor cage.  Such a cage-jump phenomenon has been proposed to be the building block of dynamical heterogeneity \cite{Candelier1,Candelier2}, aging dynamics \cite{Lee}, and long-time diffusion \cite{Pastore}.  Cage jumps happen because along some specific directions, {\it e.g.}~directions along low-frequency quasilocalized modes, the energy barriers are so low that can be overcome by the thermal energy \cite{xu,Heuer}.  Since pinning particles with the largest $\Psi$ drives the boson peak to higher frequencies and significantly weakens the low-frequency quasilocalization, we expect that the pinning leads to a dramatic increase of the cage-jump temperature $T_j$.

We follow the same protocol as in \cite{Lee} to identify cage jumps and define $T_j$ as the lowest temperature above which cage jumps occur within a simulation time of $3000$.  As shown in Fig.~\ref{fig3}(a), both pinning protocols increase $T_j$ when increasing the number of pinned particles $N_p$, which is more remarkable for the pinning of particles with the largest $\Psi$.  This suggests that cage jumps are mainly triggered by particles with the largest $\Psi$.  Figure~\ref{fig3}(b) shows the change of the $T_j$ distribution with only $3\%$ particles being pinned.  Both pinning protocols move the whole distribution to much higher temperatures.  Pinning particles with the largest $\Psi$ also shifts the maximum of the distribution from the low $T_j$ end for the no pinning case to the middle of the distribution, indicating that the glass stability is indeed enhanced.

\begin{figure}[t]
    \includegraphics[width=0.45\textwidth]{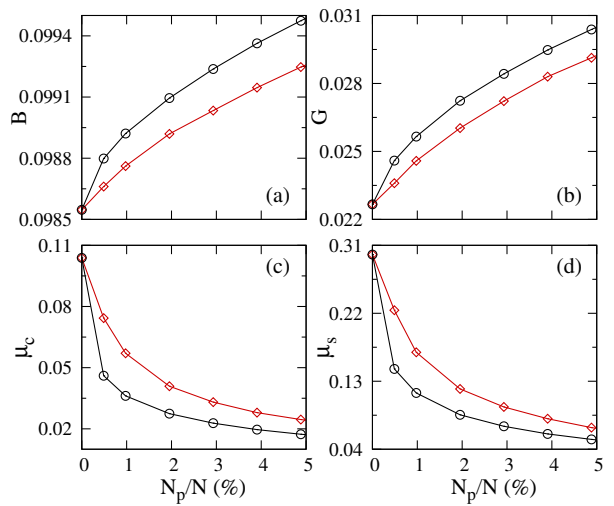}
  \caption{(color online). Dependence of (a) bulk modulus $B$, (b) shear modulus $G$, (c) nonaffinity of compression $\mu_c$, and (d) nonaffinity of shear $\mu_s$ on the fraction of pinned particles $N_p/N$ for 3D $T=0$ glasses.  The circles and diamonds are for the pinning of particles with the largest $\Psi$ and random pinning, with the solid curves to guide the eye.}
  \label{fig4}
\end{figure}

The role of particles with the largest $\Psi$ is further investigated via the mechanical response of glasses to the pinning.  When a compression or shear is imposed, all particles are initially displaced affinely.  Energy minimization is then performed without allowing the pinned particles to move.  Lees-Edwards boundary conditions are applied to mimic shearing \cite{allen}. As shown in panels (a) and (b) of Fig.~\ref{fig4}, pinning particles hardens the glass by boosting both the bulk and shear moduli, especially the shear modulus.  Again, pinning particles with the largest $\Psi$ induces more efficient hardening than random pinning.  This hardening effect contains important implications.  The elastic moduli can be decomposed into two terms, the Born term from affine deformation and a second one from nonaffine deformation \cite{Maloney}.  In the absence of nonaffine deformation, pinning particles would not cause any change of the elastic moduli.  Therefore, the increase of moduli with pinning indicates that pinning particles  significantly alters the nonaffinity, which thus reveals the strong connection between nonaffinity and structural heterogeneity.

In panels (c) and (d) of Fig.~\ref{fig4}, we show the evolution of nonaffinity upon compression and shear with the number of pinned particles.  The nonaffinity is evaluated by the ratio of nonaffine and affine particle displacements $\mu_{c,s}=\sum_i\left(\delta r^{c,s}_{i,{\rm NA}}\right)^2/\sum_i\left( \delta r^{c,s}_{i,{\rm A}}\right)^2$, where the subscripts and superscripts $c$ and $s$ denote compression and shear, $\delta r^{c,s}_{i,{\rm NA}}$ and $\delta r^{c,s}_{i,{\rm A}}$ are the nonaffine and affine displacement of particle $i$ under a strain of $5\times 10^{-6}$, and the sums are over all unpinned particles.  Both pinning protocols efficiently suppress the nonaffinity.  Strikingly, pinning only $0.5\%$ particles with the largest $\Psi$ significantly reduces the nonaffinity to half of the no pinning case.  This observation reveals that the nonaffinity is mainly originated from particles with the most defective local structures.

By pinning a small fraction of particles with the largest $\Psi$, we greatly suppress the low-frequency quasilocalization, enhance the glass stability subject to thermal excitations, harden the glass with larger elastic moduli, and weaken the nonaffine deformation.  These observations indicate that the order parameter $\Psi$ defined from the normal modes of vibration based on the energy equipartition is valid to characterize the structural heterogeneity of disordered solids.  Therefore, the low-frequency quasilocalization, instability, softening, and nonaffinity of disordered solids arise from a small number of particles with the most defective local structures.  The excellent agreement of the spatial distributions between $\Psi$ and low-temperature dynamics suggests that the Debye-Waller factor can be used as the experimental probe of the structural heterogeneity of colloidal glasses.  Pinning particles with the largest Debye-Waller factors using optical tweezers is feasible to verify our observations in experimental colloidal systems.

We are grateful to A. J. Liu and H. Tanaka for helpful discussions. This work is supported by National Natural Science Foundation of China No. 21325418 and 91027001, National Basic Research Program of China (973 Program) No. 2012CB821500, CAS 100-Talent Program No. 2030020004, and Fundamental Research Funds for the Central Universities No. 2340000034.

\end{document}